\documentclass[preprint,3p,twocolumn]{elsarticle}

\usepackage{amssymb}
\usepackage{amsmath}
\usepackage{psfrag}
\usepackage{amsopn}
\usepackage{color}

\newcommand{\br}{{\bf r}} 
 
\newcommand{\bp}{{\bf p}} 
\newcommand{\bG}{{\bf G}} 
\newcommand{\bE}{{\bf E}} 
\newcommand{\bI}{{\bf I}} 
\newcommand{\Gint}{\mathbf{G}^{\mathrm{int}}}
\newcommand{\rhoR}{\rho_{\mathrm{R}}}
\newcommand{\rhoNR}{\rho_{\mathrm{NR}}}

\journal{Photonics and Nanostructures}

\begin{document}

\begin{frontmatter}
   \title{Radiative and non-radiative local density of states \\ on disordered plasmonic films}
   \author{A. Caz\'e, R. Pierrat}
   \author{R. Carminati\corref{remi}}
   \address{Institut Langevin, ESPCI ParisTech, CNRS, 10 rue Vauquelin, 75231 Paris Cedex 05, France}
   \ead{remi.carminati@espci.fr}

   \begin{abstract}
      We present numerical calculations of the Local Density of Optical States (LDOS) in the near field of disordered plasmonic films.
      The calculations are based on an integral volume method, that takes into account polarization and retardation effects, and allows us to discriminate
      radiative and non-radiative contributions to the LDOS. At short distance, the LDOS fluctuations are dominated by non-radiative channels, showing that
      changes in the spontaneous dynamics of dipole emitters are driven by non-radiative coupling to plasmon modes. 
      Maps of radiative and non-radiative LDOS exhibit strong fluctuations, but with substantially different spatial distributions.
   \end{abstract}


\end{frontmatter}

\section{Introduction}
\label{intro}

Disordered plasmonic films obtained by evaporating noble metals on a substrate are known to exhibit unusual optical properties~\cite{shalaev}. 
Close to the percolation threshold, metallic clusters with fractal perimeters leads to the emergence of subwavelength areas supporting enhanced electric field, commonly called hot spots~\cite{gresillon99}.
These randomly distributed localized fields turned out to be very promising for sensing~\cite{moskovits85,fort08}, subwavelength focusing~\cite{li08}, or non-linear optics~\cite{sarychev00}. 
Although several theoretical and numerical works have been reported on the subject, the question of the Local Density of Optical States (LDOS) has been hardly addressed.

It has been known for long that the decay rate of a fluorescent emitter depends on its electromagnetic environment~\cite{purcell46,chance78}, the dependence being described by the LDOS $\rho(\br,\omega)$, with $\br$ the location of the emitter and $\omega$ the emission frequency. Indeed, the lifetime $\tau$ of the excited state of
a dipole emitter with transition dipole $\bp$ is given in perturbation theory by $1/\tau = \pi\omega|\bp|^2\rho(\br,\omega)/(3\epsilon_0\hbar)$ where $\epsilon_0$ is the vacuum permittivity and $\hbar$ the reduced Planck constant.
Thus the LDOS can be directly probed experimentally by measuring $\tau$. In a disordered medium, changes in the LDOS probe the local 
environment~\cite{froufe07,froufe08,sapienza11}, the
photon transport regime~\cite{beenakker02,pierrat10} or drive long-range correlations of speckle patterns~\cite{skipetrov06,caze10}.
Recently, LDOS statistics in the vicinity of disordered films have been studied experimentally~\cite{krachmalnicoff10}. Enhanced LDOS fluctuations have been observed 
close to the percolation threshold, in a regime where the film morphology is controlled by fractal clusters. These enhanced fluctuations have been qualitatively associated
to localized plasmon modes.
Theoretical and numerical studies of semi-continuous disordered metallic films are very often based on approximations, such as mean-field theories~\cite{Stroud88} or quasi-static calculations~\cite{sarychev00,stockman01}. 
An exact numerical approach has been reported recently using a FDTD (Finite-Difference Time-Domain) scheme~\cite{Shalaev10}.

In this paper, we present numerical calculations of the LDOS in the vicinity of disordered metallic films based on an integral volume method. This exact formulation is
limited only by the discretization of the films into finite size cells. The numerical algorithm is divided into two steps. Firstly,
we use a Monte-Carlo algorithm to simulate the growth of a gold film under an evaporation/deposition process, and check that the geometrical properties of the film
near the percolation threshold are in good agreement with experimental observations.
Secondly, we solve Maxwell's equations in 3D, taking into account polarization and retardation effects, which allows us to compute maps and statistical distributions of the LDOS. The computations are in agreement with known experimental results. 
The approach allows us to split the LDOS into its radiative and non-radiative contributions, and to discuss their relative contributions to the spatial fluctuations of the LDOS,
which is the main focus of this work.

\section{Numerical approach}
\label{numericalapproach}

\subsection{Generation of disordered films}
\label{filmgeneration}

Our first goal is to generate numerically disordered metallic films that have the same properties as the experimental evaporated metallic films. To do so, we use a kinetic Monte-Carlo algorithm, as proposed in~\cite{aubineauphd}. 
The idea is to randomly deposit 5-nm large gold particles on a square grid {\it via} an iterative algorithm, and let the particles diffuse under the influence of 
an interaction potential until a stable geometry is reached.
At every iteration of the algorithm, we randomly choose either to deposit a new particle (probability $p_0$) or to make a particle on the grid jump to a more stable neighbour site (probability $p_{i\rightarrow j}$ to scatter from site $i$ to site $j$).
Using the normalization $p_0 + \sum_{i,j\ne i} p_{i\rightarrow j}=1$, we only need to pick a random number out of $[0,1]$ to determine the relative weight of each
process.
More precisely, the probability to deposit a particle reads $p_0 = N F$, where $N$ is the number of particles that remains to be deposited in order to reach the prescribed filling fraction, and $F$ is a constant (with dimension $s^{-1}$) modeling the experimental deposition rate. The probability for a particle located on site $i$
to jump to the neighbour site $j$ reads $p_{i\rightarrow j} = \exp[-\Delta E_{i\rightarrow j}/(k_B T)]$, where $k_B$ is the Boltzmann constant,
$T$ the temperature of the surface and $\Delta E_{i\rightarrow j}$  the activation energy barrier.
Computing $\Delta E_{i\rightarrow j}$ is a complex issue for atoms~\cite{ferrando94,mottet98}, and is not possible from first principles for nanometer size particles.
In the present approach, we have chosen to deal with a rescaled atomic potential that renormalizes the energy barrier in order to apply to a nanoparticle. We assume that
$\Delta E_{i\rightarrow j} = \alpha (E_i - E_j)$, where $\alpha$ is a positive dimensionless adjustable parameter taking into account the influence of the substrate
and the scaling. 
$E_i$ is the rescaled ``atomic'' potential of a particle located on site $i$, which is allowed to
jump to the neighbour site $j$ if $E_i>E_j$.
This potential is given by the following expression based on a tight-binding second moment method~\cite{cleri93}:
\begin{multline}
	E_i = A\sum_{i\neq j} \exp[-p(r_{ij}/r_0 - 1)]\\
	- B\left\{\sum_{i\neq j} \exp[-2q(r_{ij}/r_0 - 1)] \right\}^{1/2}.
\end{multline}
In this expression, $r_0$ is the size of one particle, $r_{ij}$ the distance between two sites $i$ and $j$ and $A$, $B$, $p$ and $q$ are constants that were tabulated for
atoms~\cite{cleri93}. 
The iterative deposition process is stopped when all particles have been deposited (so that  the prescribed filling fraction has been reached) 
and no particle can move to a more stable site.

\begin{figure}[h!]
	\begin{center}
		\psfrag{size}{\small{$375\,\textrm{nm}$}}
		\psfrag{f20}[c]{\small{$f=20\,\%$}}
		\psfrag{f50}[c]{\small{$f=50\,\%$}}
		\psfrag{f75}[c]{\small{$f=75\,\%$}}
		\includegraphics[width=\linewidth]{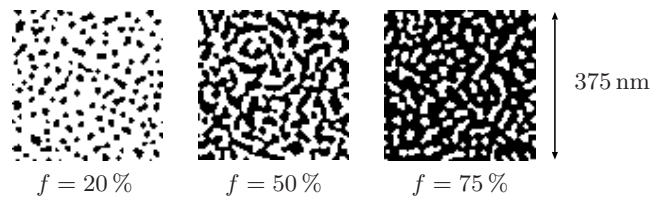}
		\caption{Numerically generated gold films for three different filling fractions $f$ (gold is represented in dark). The parameters for the computation are: $T=300 \,\mathrm{K}$, $\alpha = 2.58.10^{-2}$, $F = 10^{14}\,\mathrm{s}^{-1}$, $A = 0.2061\,\mathrm{eV}$, $B = 1.79\,\mathrm{eV}$, $p = 10.229$, $q = 4.036$.}
		\label{fig:films}
	\end{center}
\end{figure}

\begin{figure}[h!]
	\begin{center}
		\psfrag{f20}[c]{\small{$f=20\,\%$}}
		\psfrag{f50}[c]{\small{$f=50\,\%$}}
		\psfrag{P}[c]{\small{Perimeter (nm)}}
		\psfrag{S}[c]{\small{Surface (nm$^2$)}}
		\psfrag{euclidian}[c]{$S^{0.5}$}
		\psfrag{fractal}[c]{$S^{0.94}$}
		\psfrag{375}[c]{\small $375\,\textrm{nm}$}                     
		\includegraphics[width=\linewidth]{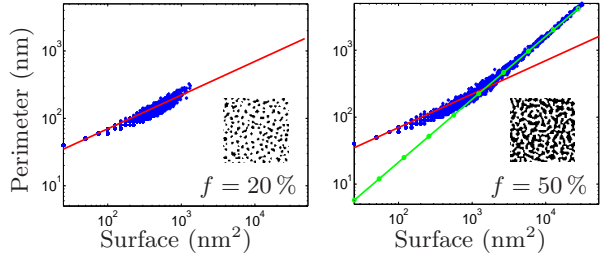}      
		\caption{Distribution in a perimeter/surface diagram of the clusters taken out from 100 numerically generated films. Left: filling fraction $f=20\,\%$.
		Right: filling fraction $f=50\,\%$. The red solid line and green dotted line are guides for the eye, corresponding to $P = 7 S^{1/2}$ and $P = 0.28 S^{1.88/2}$, respectively.} 
		\label{fig:surface_perimeter}
	\end{center}
\end{figure}

Three examples of films, with a lateral size of $375\,\mathrm{nm}$ and three different surface filling fractions $f$, are shown in Fig.~\ref{fig:films}.
When the filling fraction increases, a continuous metallic path appears linking two sides of the sample (percolation). A very important feature of the disordered metallic films is the apparition of clusters with fractal perimeter near the percolation threshold~\cite{shalaev}.
The perimeter $P$ of a cluster is said to be fractal when $P_{\mathrm{fractal}} \propto S^{D/2}$, where $S$ is the cluster surface and $D$ is a non-integer number called fractal dimension~\cite{mandelbrot}. Usual euclidian 2D surfaces have a perimeter satisfying $P_{\mathrm{euclidian}} \propto S^{1/2}$.
It has been shown experimentally that on disordered metallic films, the fractal dimension is $D=1.88$~\cite{gadennephd}. 
To check this feature, we generated 100~films with filling fractions $f=20\,\%$ and $f=50\,\%$. 
We extracted the perimeter and surface of all clusters in all numerically generated films. We show in Fig.~\ref{fig:surface_perimeter} the location of each cluster 
in a perimeter/surface diagram, in a log-log scale (each blue cross corresponds to one cluster), for both filling fractions.
For low filling-fraction, every cluster has an euclidian perimeter ($D=1$). For filling fraction $f=50\,\%$, we clearly see the existence of fractal clusters with $D\simeq 1.88$. 
This result, already shown in~\cite{aubineauphd}, is a strong evidence that the geometrical features of experimental disordered films are well described by the numerical generation method.

\subsection{Expression of the LDOS}

In order to compute the electric field and the LDOS on disordered films, we consider that a unit pixel of a numerically generated film as that shown in Fig.~\ref{fig:films} is a 5-nm size gold cube described by its dielectric constant $\epsilon(\omega)$, taken from~\cite{palick}.
To compute the LDOS $\rho(\br_0,\omega)$, we have to compute the imaginary part of the dyadic electric Green function $\bG$ at the position of the emitter~\cite{chance78,wylie84}. The normalized LDOS reads:
\begin{equation}
   \label{ldos}
   \frac{\rho}{\rho_0} = \frac{6\pi}{k_0} \operatorname{Im} \left[\operatorname{Tr} \bG(\br_0,\br_0,\omega) \right].
\end{equation}
$k_0=\omega/c=2\pi c/\lambda$ and $\rho_0=\omega^2/(\pi^2 c^3)$ is the LDOS in free space. 
The dyadic Green function $\bG$ connects an electric dipole $\bp$ at position $\br'$ to the radiated electric field at position $\br$ through the relation $\bE(\br,\omega) = \mu_0\omega^2\bG(\br,\br',\omega)\bp$. It describes the electromagnetic response of the environment.

\subsection{Calculation of the dyadic Green function}

To compute the dyadic Green function $\bG$ in the presence of the film, we consider an electric dipole $\bp$ located at position $\br_0$, and use a volume integral method~\cite{harrington}. The electric field at any point $\br$ obeys the volume integral equation (Lippmann-Schwinger equation)
\begin{multline}
		\label{integral_formula}
		\bE(\br,\omega) = \mu_0\omega^2\bG_0(\br,\br_0,\omega)\bp \\
		+ k_0^2[\epsilon(\omega)-1]\int_V \bG_0(\br,\br',\omega)\bE(\br',\omega)\mathrm{d}^3\br'
\end{multline}
where $V$ is the volume occupied by the metallic film. $\bG_0$ is the dyadic Green function of free space, given by~\cite{vanbladel91,yaghjian80}
\begin{multline}
	\bG_0(\br,\br',\omega) = \operatorname{PV}\left[\bI + \frac{\nabla\nabla}{k_0^2}\right]\frac{\exp(ik_0 R)}{4\pi R} \\
	- \delta(\br - \br') \frac{\bI}{3k_0^2},
\end{multline}
where $R = |\br - \br'|$, $\operatorname{PV}$
denotes the Principal Value operator and $\delta$ is the Dirac delta function.
In order to solve the integral equation numerically, we discretize $V$ into cells of size $\Delta$, and assume that the electric field is constant in each cell (the volume of cell number $j$ will be denoted by $V_j$).
For all calculations presented in this paper, $\Delta$ is set to $2.5\,\mathrm{nm}$ so that each gold cube is divided into eight cells.
To improve convergence of the numerical computation, we integrate the Green dyadic on the cell volume (moment method)
and define $\Gint_{ij} = \int_{V_j} \bG_0(\br_i,\br',\omega)\,\mathrm{d}^3\br'$.
To calculate the electric field in each cell, we have to solve the following linear system:
\begin{multline}
		\label{system}
		\left\{ \bI - k_0^2\left[\epsilon(\omega)-1\right]\Gint_{ii}\right\} \bE_i - k_0^2\left[\epsilon(\omega)-1\right]\sum_{j\neq i} \Gint_{ij}\bE_j \\
		= \mu_0\omega^2\bG_0(\br_i,\br_0,\omega)\bp.
\end{multline}
The solution leads to the expression of the three components of the electric field $\bE_i $ in cell number $i$, for all $i$.
The computation of $\Gint_{ii}$ has to be performed with care, due to the singularity of the Green function $\bG_0$ at the origin.
This can be done in Fourier space, using the Weyl expansion as exposed in~\cite{chaumet04}.
Solving Eqs.~(\ref{system}) for three orthogonal orientations of the source dipole gives direct access to the full dyadic Green function $\bG(\br,\br_0,\omega)$.
The LDOS is deduced from Eq.~(\ref{ldos}) (note that the imaginary part of the Green dyadic is not singular at the origin for $\br_0$ in vacuum).

The numerical approach also allows us to calculate separately the radiative LDOS $\rhoR$ (which is proportional to the far-field power radiated by the dipole source) and the non-radiative LDOS $\rhoNR$ (which is proportional to the power absorbed inside the metal)~\cite{bian95,klimov01}. Energy conservation requires that $\rho = \rhoR + \rhoNR$, so that only two quantities need to be calculated. We can compute the normalized non-radiative LDOS from
\begin{equation}
   \label{ldos_nr}
   \frac{\rhoNR}{\rho_0} = \frac{6\pi\epsilon_0^2}{k_0^3|\bp|^2}\operatorname{Im}[\epsilon(\omega)]\int_V |\bE(\br',\omega)|^2\,\mathrm{d}^3\br'
\end{equation}
and then deduce $\rhoR$ by subtraction. Equation~(\ref{ldos_nr}) is discretized the same way as Eq.~(\ref{integral_formula}).
>From such calculations it is possible to address the contribution of radiative and non-radiative modes to the LDOS, as we will see. This is an important issue
in the understanding of the optical response of disordered metallic films, and their use for the control of the dynamics of fluorescent sources.

\section{Results}
\label{results}

\subsection{Mapping the LDOS and its radiative and non-radiative components}
\label{results_maps}

\begin{figure}[h!]
	\begin{center}
		\psfrag{f20}[c]{\footnotesize{$f=20\,\%$}}
		\psfrag{f50}[c]{\footnotesize{$f=50\,\%$}}
		\psfrag{375}[c]{\footnotesize{$375\,\mathrm{nm}$}}
		\psfrag{gamma}[c]{\small{$\rho/\rho_0$}}
		\psfrag{gammanr}[c]{\small{$\rhoNR/\rho_0$}}
		\psfrag{gammar}[c]{\small{$\rhoR/\rho_0$}}
		\includegraphics[width=\linewidth]{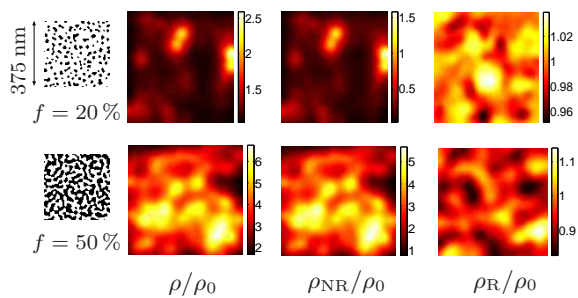}
		\caption{Maps of the total ($\rho$), non-radiative ($\rhoNR$) and radiative LDOS ($\rhoR$) normalized by the LDOS in vacuum ($\rho_0$) at $40\,\textrm{nm}$ distance above two films with filling fractions $f=20\,\%$ and $f=50\,\%$. The wavelength is $\lambda = 780\,\mathrm{nm}$. Note that the color scale is different for every map.}
		\label{fig:maps}
	\end{center}
\end{figure}

Using the approach described in section~\ref{numericalapproach}, we computed maps of the total, radiative and non-radiative LDOS at 40 nm above numerically generated disordered metallic films. This distance has been chosen since it provides substantial near-field effects and remains compatible with standard computational resources. A full study of the 
distance dependence, both theoretical and experimental, will be published elsewhere~\cite{castanie12}.
The results are shown in Fig.~\ref{fig:maps}.
We clearly see that near the percolation threshold (film with $f=50\,\%$), complicated LDOS structures appear, with local enhancements on subwavelength areas. The existence of local enhancements of the electric field intensity (hot spots) is a well-known result, that was observed before in experiments~\cite{gresillon99}. These local field
enhancements directly translate into LDOS enhancements, leading to strongly fluctuating LDOS patterns.
Another interesting output of the calculations is that at a distance $40\,\mathrm{nm}$ above a film with $375\,\textrm{nm}$ lateral extension, LDOS spatial
fluctuations are mainly due to non-radiative channels (this can be seen by comparing the standard deviations of $\rhoNR$ and $\rhoR$ in Fig.~\ref{fig:maps}).
Moreover, the spatial distribution of the radiative LDOS $\rhoR$ is completely different from that of the non-radiative LDOS $\rhoNR$. In a fluorescence experiment using single nanoscale emitters, this means that the trade-off between radiative and non-radiative decay is dependent on the emitter position. The apparent 
quantum yield also becomes a spatially fluctuating quantity, with expected strong fluctuations.

\subsection{Statistical distributions of $\rho$, $\rhoNR$ and $\rhoR$}
\label{results_histograms}

\begin{figure}[h!]
	\begin{center}
		\psfrag{gamma}{\small{$\rho/\rho_0$}}
		\psfrag{gammanr}{\small{$\rhoNR/\rho_0$}}
		\psfrag{gammar}{\small{$\rhoR/\rho_0$}}
		\psfrag{occurences}{\small{Occurences}}
		\psfrag{f20}[c]{\small{$f=20\,\%$}}
		\psfrag{f50}[c]{\small{$f=50\,\%$}}
		\includegraphics[width=0.8\linewidth]{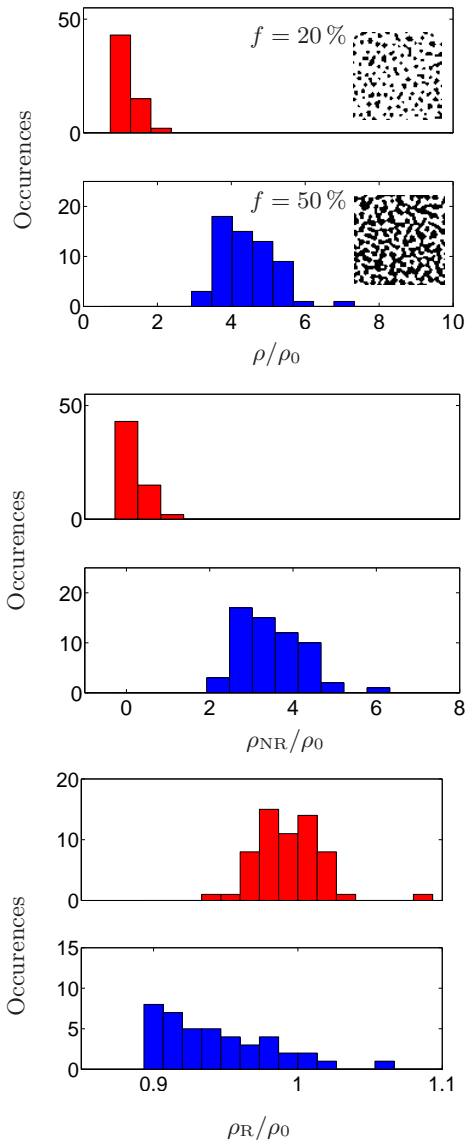}
		\caption{Histograms of the total ($\rho$), non-radiative ($\rhoNR$) and radiative LDOS ($\rhoR$) normalized by the LDOS in vacuum ($\rho_0$) at $40\,\textrm{nm}$ distance above two series of films of same filling fraction (red: $f=20\,\%$; blue: $f=50\,\%$). Every generated film has a lateral size of $375\,\mathrm{nm}.$}
		\label{fig:histograms}
	\end{center}
\end{figure}

The existence of localized modes on disordered metallic films was recently studied experimentally measuring the statistical distribution of the LDOS~\cite{krachmalnicoff10}. It was shown that the apparition of fractal clusters was correlated to enhanced fluctuations of the LDOS, that are a direct signature of the presence of spatially localized
field distributions.
We computed the statistical distribution of the total, non-radiative and radiative LDOS, for two collections of films of lateral size $375\,\mathrm{nm}$ with filling fractions $f=20\,\%$ and $f=50\,\%$. For each filling fraction, we generated 60~different films and computed the value of the LDOS at a distance $40\,\mathrm{nm}$ above the center of the film. The histograms are shown in Fig.~\ref{fig:histograms}. 
>From the calculations, we recover the enhanced fluctuations of the LDOS observed in~\cite{krachmalnicoff10} close to the percolation threshold.
A comparison of the histograms for $\rho$, $\rhoNR$ and $\rhoR$ also confirms that at a distance  $40\,\textrm{nm}$ from the film, the LDOS fluctuations are mainly
driven by non-radiative channels, as already discussed in section~\ref{results_maps}.
Finally we note that the computations are performed on samples with lateral size on the order of $\lambda/2$, so that the LDOS spatial distribution might be affected
by finite-size effects. Although not shown for brevity, we have performed computations with sample sizes from $150 \,\textrm{nm}$  to $375\,\textrm{nm}$. 
These computations have shown that although the statistical distribution of $\rhoR$ is size-dependent in this regime, the distribution of $\rho$ and $\rhoNR$ are quite robust.

\subsection{Correlation between LDOS hot spots and film topography}
\label{maps_topography}

\begin{figure}[h!]
	\begin{center}
		\psfrag{375}[c]{\small{$375\,\mathrm{nm}$}}
		\includegraphics[width=\linewidth]{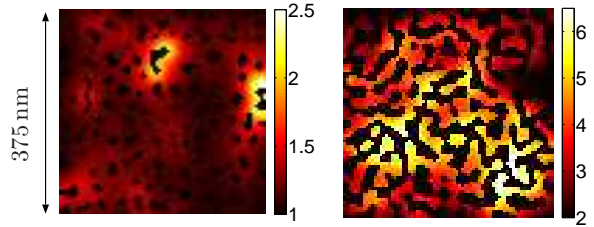}
		\caption{Maps of the total normalized LDOS at a distance $40\,\textrm{nm}$ represented on top of the film topography (gold is represented with black color). Wavelength  $\lambda = 780\,\mathrm{nm}$. Left: $f=20\,\%$. Right: $f=50\,\%$.}
		\label{fig:maps_with_film}
	\end{center}
\end{figure}

To get more insight about the origin of the localized LDOS (or intensity) enhancements, we superimpose the maps of the total normalized LDOS and the topography of the films,
as shown in Fig.~\ref{fig:maps_with_film}.
The maps clearly show that at low filling fraction (left), classical plasmon resonances of isolated particles are responsible for local enhancements of the LDOS. 
Near the percolation threshold (right), the origin of the LDOS structure is more complex. The non-trivial relation between the topography and the location of localized field enhancements is sustained by collective interactions. Finding a simple model to understand this connection is still an open issue.       

\section{Conclusion}
\label{conclusion}

In conclusion, we have presented exact 3D numerical calculations of maps and statistical distributions of the LDOS in the near-field of disordered plasmonic films.
The calculations describe the well-known existence of localized enhancements of the near-field intensity and the LDOS on subwavelength areas, for
filling fractions close to the percolation threshold.  
The method also permits a calculation of the radiative and non-radiative contributions to the LDOS. 
We have shown that at a distance $40\,\textrm{nm}$ above the film (near-field zone), the LDOS fluctuations are chiefly driven by non-radiative channels.
Nevertheless, both radiative and non-radiative LDOS exhibit strong spatial fluctuations, with completely different spatial distributions.
Understanding the trade-off between radiative and non-radiative channels is a key issue for the understanding of the optical properties of disordered plasmonic films,
and their use as sensors, absorbers or new materials for the control of light emission.

\bibliographystyle{elsarticle-num}

\end{document}